\documentclass[12pt,preprint]{aastex}

\usepackage{graphicx,amssymb}
\usepackage[usenames,dvips]{color}
\newcommand{\kms}{\ifmmode {\rm km\,s}^{-1} \else km\,s$^{-1}$\fi}

\shorttitle{Zeeman effect in the 36 GHz methanol maser in DR21W}
\shortauthors{Momjian, Sjouwerman \& Fish}

\begin{document}

\title{Very Large Array Detection of the 36 GH\lowercase{z} Zeeman Effect in
DR21W Revisited}

\author{Emmanuel~Momjian, Lor\'ant~O.~Sjouwerman} 
\affil{National Radio Astronomy Observatory,
  P.O. Box O, 1003 Lopezville Rd., Socorro, NM 87801}
\email{emomjian@nrao.edu}
\and
\author{Vincent~L.~Fish}
\affil{MIT Haystack Observatory, Route 40, Westford, MA 01886}

\begin{abstract}
  We report on the observation of the 36 GHz methanol maser line in the star
  forming region DR21W to accurately measure the Zeeman effect.
  The reported Zeeman signature by \citet{fish11} became
  suspicious after an instrumental effect was discovered in the early
  days of the Very Large Array Wide-band Digital Architecture (WIDAR)
  correlator commissioning. We conclude that the previously reported magnetic
  field strength of 58~mG\,(1.7 Hz~mG$^{-1}/z$) is
  instrumental in nature and thus incorrect. With the improved
  performance of the array, we now deduce a 3$\sigma$ limit of
  $-$4.7 to +0.4~mG\,(1.7 Hz~mG$^{-1}/z$) for the line-of-sight component of
  the magnetic field strength in DR21W.
\end{abstract}


\section{Introduction}
\label{intro}

Studies of the Class\,I methanol masers found in shock-excited
environments, e.g., in star-forming regions in the Galaxy, have
recently received a boost in interest. This interest is due to the
increased frequency coverage of radio interferometers, allowing for
higher angular resolution observations of the 
low-level excited lines, e.g., lines at 25, 36 and 44 GHz
\citep[e.g.][]{voronkov05,voronkov06,sjouwerman10}.

The brightness of a few of these masers allow the measurement of the
Zeeman effect, and thus the determination of the magnetic field of the
environment in which they occur. As a result of the large uncertainty
in the Zeeman splitting factor for methanol \citep{vlemmings11},
magnetic field strengths are commonly reported in units of the
(unknown) splitting coefficient (i.e., with $z$ = 1.7~Hz\,mG$^{-1}$
for the 36 GHz transition). Despite this uncertainty, significant results can be
deduced for magnetic fields in units of the actual value, i.e., in
terms of 1.7 Hz\,mG$^{-1}$\,/\,$z$ for the 36 GHz transition. Examples of Zeeman splitting
studies in Class\,I methanol masers are \citet{sarma09,sarma11,fish11}
and \citet[][in prep.]{momjian12}.

In 2010, the commissioning of the new, Wide-band Digital Architecture
(WIDAR) correlator of the Karl G. Jansky Very Large Array (VLA) had started \citep{perley11}.
During commissioning, the VLA continued to obtain
scientific observations with this new correlator. Together with, e.g.,
the new 26.5--40 GHz Ka-band receivers, it demonstrated the excellent
performance of the VLA with the detection of many new 36 GHz Class\,I
methanol masers \citep[e.g.][]{sjouwerman10}. As a result, a
Zeeman-like signature was observed at 36 GHz in DR21W \citep{fish11},
which ostensibly implied a line of sight magnetic field strength of
$B \cos\theta$ = 58~mG (1.7~Hz\,mG$^{-1}/z$).
This value is much larger than expected at methanol maser sites in
star-forming regions.  Indeed the authors did caution against
interpreting the Stokes\,$V$ signature as a very strong magnetic
field, but the detection looked convincing enough to warrant further
discussion in their paper. However, soon after the first publication
\citep{fish11}, the WIDAR commissioning team discovered a small but
significant unanticipated spectral response, or ``spectral splatter''
effect, while observing strong spectral lines at high frequency resolutions.
Zeeman effect studies, which in this case rely on the difference of two strong signals,
were particularly vulnerable.
This rendered the so-called S-curve profiles in the
Stokes\,$V$ spectra very unreliable \citep{sault10,sault12a,sault12b}.

Since then, and up until the problem was fully understood, all Zeeman
effect observations with the VLA were put on hold. Zeeman effect
commissioning tests were concluded in 2012 March, and in 2012
April these observations were officially
re-introduced\footnote{Commissioning tests show that observations since 2011
August could be used for spectral line polarization measurements.}.  Here we report on
the re-observation of the Zeeman effect in DR21W with the WIDAR
correlator and comment on the previously reported, uncomfortably high
value of the magnetic field strength as deduced in the 36 GHz line by
\citet{fish11}.

\section{Observations and Data Reduction}

Observations of the $4_{-1} - 3_{0}E$ methanol maser emission line at
36.169 GHz toward the star forming region DR21W were carried out with
the VLA on 2012 May 1 in two 2-hour observing sessions.  The
array was in CnB configuration, providing a synthesized beam of
$0\farcs51 \times 0\farcs40$.  The observations were performed
in dual polarization with 2 MHz bandwidth using 256 spectral channels,
resulting in a channel separation of 7.8 kHz, or a velocity separation
of 65 m\,s$^{-1}$. The total on-source time was 170 minutes after
combining both observing runs. The calibrator sources 3C286 and 3C84
were used to set the absolute flux density scale.

The editing, calibration, deconvolution and imaging of the data were
carried out using NRAO's Astronomical Image Processing System
(AIPS; \citealt{greisen03}). After applying the amplitude gain corrections of the
calibrators 3C286 and 3C48 on the target source DR21W, the data were
Doppler corrected and the spectral channel with the brightest maser
emission signal was split off. This single-channel signal was
self-calibrated in both phase and amplitude and imaged in a succession
of iterative cycles. The final phase and amplitude solutions were then
applied to the full spectral-line visibility data of the target
source, and Stokes $I$=(RCP+LCP)/2 and $V$=(RCP$-$LCP)/2 image cubes
were constructed. Further processing of the data was done using the
{\sl MIRIAD} data reduction package to remove the scaled replica of
the Stokes\,$I$ spectrum from the observed Stokes\,$V$ spectrum
through least square fitting using the equation: $$V_{\rm obs} = aI +
\frac{zB \cos\theta}{2}\frac{dI}{d\nu}\ .$$ The fit parameter $a$ is
usually the result of small calibration errors in RCP versus LCP and
is expected to be small. In these observations, $a$ is of order
10$^{-3}$.

\begin{figure}[t]
\begin{center}
\resizebox{\columnwidth}{!}{\includegraphics{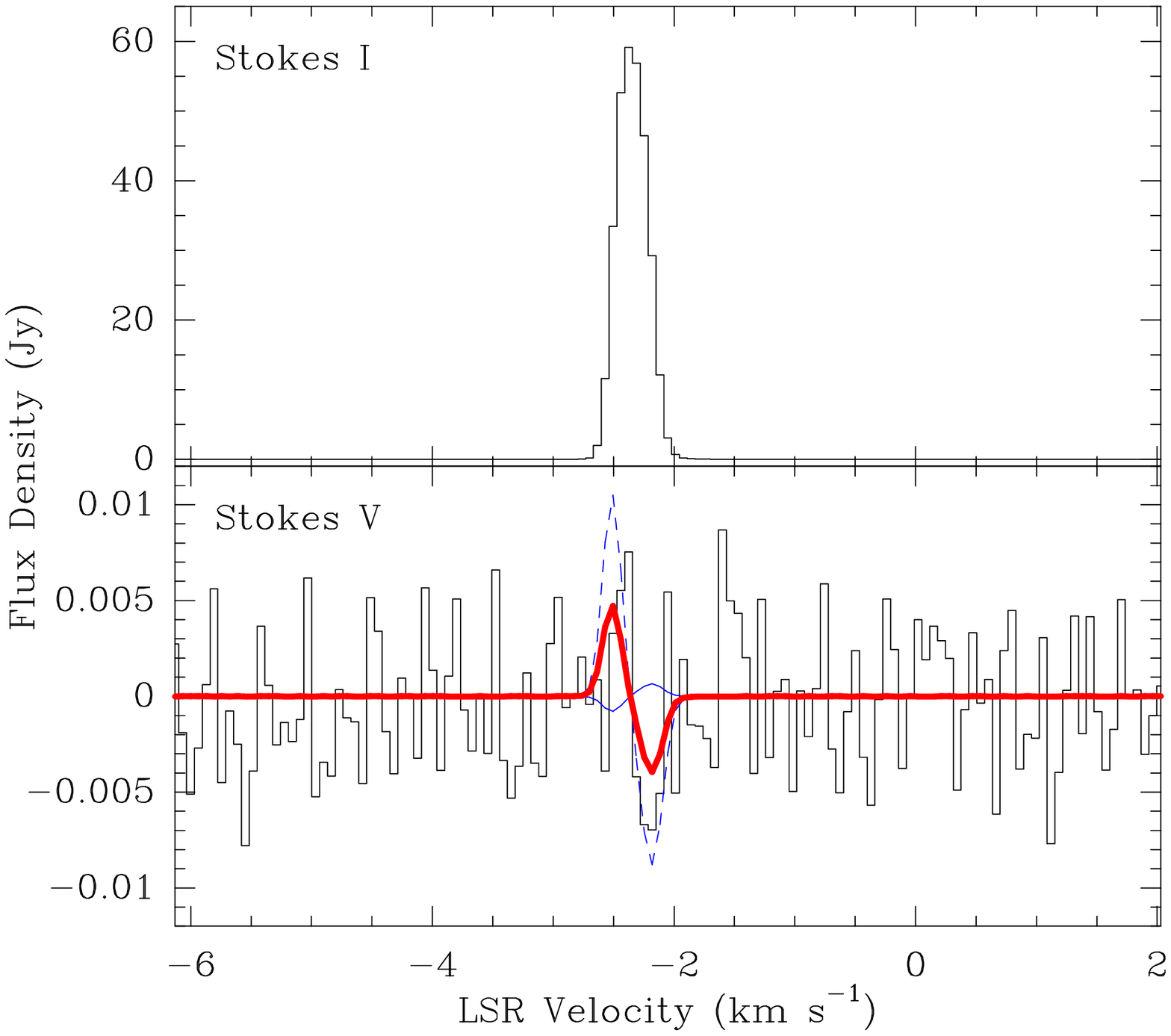}}
\caption{For a putative S-curve Zeeman effect, the thick red curve
  shows the best fit value of $-$2.1~mG~(1.7 Hz\,mG$^{-1}$\,/\,$z$) to
  the Stokes\,$V$ spectrum.  The fit shows that there is no
  significant S-curve Zeeman effect in the data, unlike the
  (instrumental) result in Figure~3 of \citet{fish11}.  The thin
  dashed and solid blue curves show the extent of the 3$\sigma$ range
  of a possible magnetic field of up to
  $-$4.7 and +0.4~mG~(1.7~Hz\,mG$^{-1}/z$), respectively.}
\label{fig1}
\end{center}
\end{figure}

\section{Results and Discussion}

Figure~\ref{fig1} shows the Stokes\,$I$ and Stokes\,$V$ spectra toward
the strongest 36 GHz methanol maser in DR21W; the same maser as in
Figure~3 in \citet{fish11}. Here, the thick red curve shows the best
fit value of $-$2.1~mG~(1.7 Hz\,mG$^{-1}$\,/\,$z$) to a putative
S-curve Zeeman signature in the Stokes\,$V$ spectrum. This best fit is
\emph{insignificant} as it is well within the noise of the Stokes\,$V$
spectrum. We note that the Zeeman splitting coefficient ($z$) of the
36 GHz methanol maser transition is currently unknown
\citep{vlemmings11}, but for proper comparison with the results of
\citet{fish11}, and \citet{sarma09}, we are quoting the values with
respect to 1.7 Hz\,mG$^{-1}$\,/\,$z$, where 1.7 Hz\,mG$^{-1}$ was
tentatively assumed as the Zeeman splitting coefficient for this maser
transition. Performing a one-dimensional $\chi^2$ fit for $V$, we
obtain a $3\sigma$ range of $-$4.7 to +0.4 mG~(1.7~Hz\,mG$^{-1}$\,/\,$z$),
depending whether the line-of-sight component
($B \cos\theta$) of the field is oriented toward ($-$) or away (+)
from the observer. These values are shown in the dashed (+$3\sigma$)
and solid ($-3\sigma$) thin blue curves in Figure~\ref{fig1}.

The results obtained from these observations show that the Zeeman
effect signature in the 36 GHz methanol maser line reported by
\citet{fish11} is not correct. Because our observations were carried
out in two observing runs, we have been able to verify that our
resulting non-detection of the Zeeman splitting in the strongest 36
GHz maser in DR21W holds for both sessions separately.

\acknowledgements

We wish to exclaim our appreciation of the
concerted effort performed by the VLA WIDAR commissioning team to
resolve the issues that led to the contamination and corruption of
Zeeman effect observations at the VLA.  The National Radio
Astronomy Observatory (NRAO) is a facility of the National Science
Foundation operated under cooperative agreement by Associated
Universities, Inc.

{\it Facilities:} \facility{VLA}.


\begin{thebibliography}{}

\bibitem[Fish et al.(2011)]{fish11}Fish, V.L., Muehlbrad, T.C.,
  Pratap, P., Sjouwerman, L.O., Strelnitski, V., Pihlstr\"om, Y.M., \&
  Bourke, T.L., 2011, \apj, 729, 14

\bibitem[Greisen(2003)]{greisen03} Greisen, E.~W.\ 2003, 
Information Handling in Astronomy - Historical Vistas, 285, 109 


\bibitem[Momjian \& Sarma(2012)]{momjian12} Momjian, E., \& Sarma,
  A.P.\ 2012, (in preparation)


\bibitem[Perley et al.(2011)]{perley11} Perley, R.A., Chandler,
  C.J., Butler, B.J., \& Wrobel, J.M.\ 2011, \apjl, 739, L1



\bibitem[Sarma \& Momjian(2009)]{sarma09} Sarma, A.P., \& Momjian, E.\ 2009,
  \apjl, 705, L176

\bibitem[Sarma \& Momjian(2011)]{sarma11} Sarma, A.P., \& Momjian, E.\ 2011,
  \apj, 730, 5

\bibitem[Sault(2010)]{sault10} Sault, R.J. 2010, EVLA Memo 148
  (http://www.aoc.nrao.edu/evla/memolist.shtml)

\bibitem[Sault(2012a)]{sault12a} Sault, R.J. 2012a, EVLA Memo 157
  (http://www.aoc.nrao.edu/evla/memolist.shtml)

\bibitem[Sault(2012b)]{sault12b} Sault, R.J. 2012b, EVLA Memo 160
  (http://www.aoc.nrao.edu/evla/memolist.shtml)

\bibitem[Sjouwerman, Pihlstr\"om \&
  Fish(2010)]{sjouwerman10}Sjouwerman, L.O., Pihlstr\"om, Y.M., \&
  Fish, V.L.\ 2010, \apjl, 710, L111

\bibitem[Vlemmings, Torres \& Dodson et al.(2011)]{vlemmings11} Vlemmings, W.H.T.,
  Torres, R.M., Dodson, R.\ 2011, A\&A, 529, 95

\bibitem[Voronkov et al.(2005)]{voronkov05} Voronkov, M.A., Sobolev,
  A.M., Ellingsen, S.P., \& Ostrovskii, A.B.\ 2005, \mnras, 362, 995

\bibitem[Voronkov et al.(2006)]{voronkov06}Voronkov, M.A., Brooks,
  K.J., Sobolev, A.M., Ellingsen, S.P., Ostrovskii, A.B., \& Caswell,
  J.L., 2006, \mnras, 373, 411

\end{thebibliography}
\end{document}